# Crack detection in beam structures with a novel Laplace based BSWI FE method

Shuaifang Zhang[1,2], Dongsheng Li[1], Wei Shen[1], Xiwen Zhang[3], Yu Liu[4]

1, Department of Civil Engineering/Engineering Mechanics, Dalian University of Technology, Dalian, China, 116023
2, Department of Mechanical Engineering, University of Florida, Gainesville, FL, USA, 23607
3, Department of Civil Engineering and architecture, University of Jinan, Jinan, Shandong, P.R.China, 250022
4, Department of Mechanical Engineering, Jiangsu Normal University, Xuzhou, China, 221100

**Abstract:** Beam structure is one of the most widely used structures in mechanical engineering and civil engineering. Ultrasonic guided wave based crack identification is one of the most important and accepted approaches applied to detect unseen small flaws in structures. Numerical simulations of ultrasonic guided wave propagation have caught more and more attention due to the fast development of hardware and software in the last few years. From all the numerical simulation methods, wavelet based finite element method has been proved to be one of the most efficient methods due to its better spatial resolution, which means it needs fewer elements to get the same accuracy and it can improve the calculation cost significantly. However, it needs a very small time interval. Laplace transform can easily convert the time domain into a frequency domain and then revert it back to a time domain. Laplace transform has thus the advantage of finding better results with a very large time interval. which can save a lot of time cost. This paper will present an innovative method combining Laplace transform and the B-spline wavelet on interval (BSWI) finite element method. This novel method allows to get results with the same accuracy and with a significantly lower time cost, which would not only decrease the total number of elements in the structure but also increase the time integration interval. The numerical Laplace transform and BSWI finite element will be introduced. Moreover, this innovative method is applied to simulate the ultrasonic wave propagation in a beam structure in different materials. Numerical examples for crack identification in beam structures have been studied for verification.

## 1 INTRODUCTION

Structures must fit the basic requirements, such as safety, durability and sustainability for a long-term operation. During such operations, their performance may be slowly weakned throughout time. It also may suffer damages or even be destroyed under serious natural disasters. Structure design and damage detection have caught a lot of attention in aerospace, mechanical structures and civil engineering {Cao, 2017; Cheng, 2017; Song, 2006; Su, 2009; Song, 2010; Zhang, 2017} and have been proved to be effective. Various approaches have been developed for damage detection in structures, such as vibration-based damage detection approaches{Yam, 2003;Chen, 2007 ;Bagheri, 2009}, the test method{An, 2015}and others.

Wavelet transform has been applied in civil engineering for a long time in many areas due to its multi-resolution capability. The Wavelet based finite element method is a relatively new numerical simulation. Wavelet is a good tool for both time and frequency analysis {Zhou, 2003}. One of the methods used in a frequency domain called the wavelet spectral finite element method{Mitra, 2005}, while the other one used in a time domain is usually called wavelet based finite element method{Xiang, 2007; Amiri, 2015}. The WSFEM is used in wave propagation in structures and can determine the exact solution but is still not ideal for damage detection. On the other hand, the WFEM is very to the traditional finite element method but it provides a better way for mesh refinement and has been applied in damage detection. It has two prominent advantages{Xiang, 2013}, one of which is that the scale is directly upgraded via the scaling functions to form multi-scale approximation bases. The other advantage is that the nesting approximation is obtained via the relationship between scaling bases and wavelet spaces. Many WFEM models have been developed and applied in nondestructive testing in different structures{Ma, 2003}. B-splined wavelet finite element models{Chen, 2012} are the most popular ones. The WFEM can be easily applied to damage detection and has a strong advantage of mesh refinement: only very few elements are needed for accurate analysis. However, the time step is still required to be very small, so the computational cost is still high.

In this paper, a novel method that combines numerical Laplace transform with BSWI FEM is proposed to simulate the wave propagation in bar structures. This novel Laplace based BSWI FEM (LBSWI) will combine the advantages of BSWI FEM and Laplace transform, which not only has high spatial resolution with a better mesh but also avoids the periodic assumption of FFT. Thus, Laplcae based BSWI FEM combines the strength of WFEM which builds a highly accurate finite element model with the strength of Laplace transform, which will transform the time domain to a frequency domain for faster calculations. Hence, LBSWI can be applied to wave propagation as well as ultrasonic damage detection with very good accuracy and low computational cost. Numerical simulations of ultrasonic wave propagation in a beam structure will be carried out. The results are compared with other different methods such as traditional FEM, WFEM, SEM. The influence of different material properties will also be studied. Besides, numerical simulations of damage detection in beam structures will be done using this novel method.

## 2 LAPLACE BASED BSWI IN BEAM STRUCTURES
### 2.1 B-spline wavelet on interval finite element formulation

Wavelet transform is based on two different types of functions, one is the wavelet function while the other is the scaling function. Both the scaling function $\phi(x)$ and wavelet funciton $\varphi(x)$ satisfy the following two scaling equations:

$$\phi(x) = \sum_{i=0}^{N-1} p_i \phi(2x - i) \quad (1)$$

$$\varphi(x) = \sum_{i=2-N}^{1} (-1)^i p_{1-i} \varphi(2x - i) \quad (2)$$

Where N is an integer and $p_i$ $(i = 0,1,\dots,N-1)$ are filter coefficients.

The shape function built in the BSWI finte element method is mainly based on organizing the scaling functions. B-spline wavelet function is built with a piecewise polynomial by joining different knots together on the interval. In order to have at least one inner wavelet on the interval [0,1], for any picked scale $j$, the dimension of the $m$th order B-spline scaling function must satisfy the following equation:

$$2^j \geq 2m - 1 \quad (3)$$

0 scale $m$th order B-spline scaling function and the wavelet function are developed by Goswami{Goswami, 1995}. All the scaling functions $\emptyset_{m,k}^j(\xi)$ could be derived by the following equations:

$$\emptyset_{m,k}^j(\xi) = \begin{cases} \emptyset_{m,k}^l(2^{j-l}\xi), k = -m+1, \cdots, -1 & (0\ boundary) \\ \emptyset_{m,2^j-m-k}^l(1 - 2^{j-l}\xi), k = 2^j - m - 1, \cdots, 2^j - -1 & (1\ boundary) \\ \emptyset_{m,0}^l(2^{j-l}\xi - 2^{-l}k), k = 0, \cdots, 2^j - m & (inner) \end{cases} \quad (4)$$

In this paper, 11 basic equations of the 4$^{th}$ scale 3$^{rd}$ order scaling functions are picked to build the B-spline wavelet on interval finite element, the scaling function plots are shown in Fig. 1.

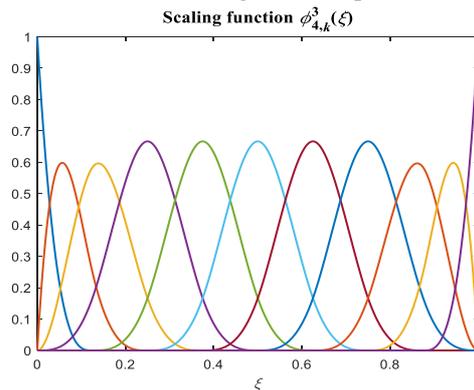

**Fig. 1** The plots of eleven BSWI4,3 scaling functions $\emptyset_{4,k}^3(\xi)(k = -3, \cdots, 7)$

The governing equations for ultrasonic wave propagation in isotropic structure is based on Navier equations:

$$(\lambda + \mu)\frac{\partial \theta}{\partial x_i} + \mu \nabla^2 u_i + \rho f_i = \rho \frac{\partial^2 u_i}{\partial t^2} \quad (5)$$

where $\lambda, \mu$ are Lame constants, $\theta = \frac{\partial u_1}{\partial x_1} + \frac{\partial u_2}{\partial x_2} + \frac{\partial u_3}{\partial x_3}$, $\rho$ represents the density of the structure, $f$ is body force, $u_i$ is the displacement along the $i$th direction.

For one dimensional classical rod structures, by transforming any subdomain [$a,b$] to basic BSWI wavelet subdomain [$0,1$], where the basic 1D element is shown below,

# Crack detection in beam structure with a novel Laplace based BSWI FE method

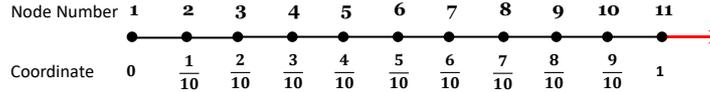

**Fig.2.** Basic BSWI43 1D element DOFs

The displacement as a function of scaling function and wavelet coefficients can be expressed as,
$$u(\xi) = \sum_{k=-m+1}^{2^j-1} a_{m,k}^j \phi_{m,k}^j(\xi) = \Phi a^e \quad (6)$$

Where $\Phi = \left[\phi_{m,-m+1}^j(\xi), \cdots \phi_{m,2^j-1}^j(\xi)\right]$ is a vector that consists of BSWI43 scaling functions, $a^e = \left[a_{m,-m+1}^j(\xi) \cdots a_{m,2^j-1}^j(\xi)\right]^T$ is the wavelet interpolation coefficient vector. FEM is based on the nodal displacement so that the new displacement equation must be established. By assigning the variant $\xi$ as each node's coordinates, we find the nodal displacement vector
$$u^e = \Phi a^e = [R^e]^{-1} a^e \quad (7)$$

Where $u^e = [u(\xi_1)\ u(\xi_2)\ \cdots\ u(\xi_n)]^T$ is the vector on each node of nodal DOFs, $[R^e]^{-1} = [\Phi^T(\xi_1)\ \Phi^T(\xi_2)\ \cdots\ \Phi^T(\xi_{n+1})]$, by substituting the solution of $a^e$ in Eq. (7) into Eq.(6), we could get the displacement as a function of nodal DOFs,
$$u(\xi) = \Phi R^e u^e = \Phi R^e u^e = N^e u^e \quad (8)$$

In which $N^e = \Phi R^e$ is the shape function we will use in the BSWI finite element model. For covenience, the plots for the shape functions are shown in Fig. 3, because there are 11 shape functions so we will separate the plots into two parts, one of which is shape functions from N1 to N6, and the other one plots shape functions from N7 to N11.

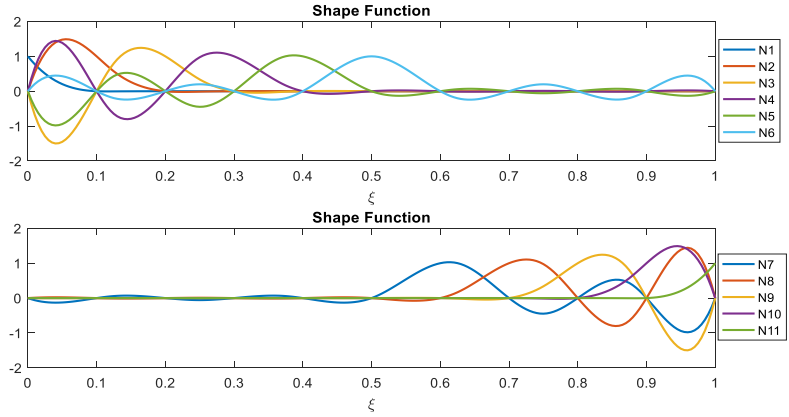

**Fig.3.** Plots of BSWI43 1D element shape functions

Apparently, the shape functions satisfy the fundamental property as described in the finite element methods:
$$N_i(\xi_j) = \delta_{ij} \quad (9)$$

Where $\delta_{ij}$ is the Kronecker delta function. From the plots of shape functions we see that they are slightly different from the shape functions in traditional FEM, in which the shape function values are located in [0,1]. Then we can assemble all the finite element matrices such as the mass matrix and stiffness matrix following the basic procedures in FEM as shown in equation 10 and 11
$$M^e = \int (N^e)^T \mu^e N^e dx \quad (10)$$
$$K^e = \int (B^e)^T D^e B^e dx \quad (11)$$

Where $\mu^e$ is the mass density per unit length, $B^e$ is the gradient matrix of shape functions, $D^e$ is the material stiffness matrix. By assembling these element matrices, we could find the final Finite Element model for ultrasonic wave propagation without damping:
$$M\ddot{U} + KU = F(t) \quad (12)$$

Where **M** is the global mass matrix, **K** is the global stiffness matrix, $F(t)$ is the time domain excitation force vector, $u(t)$ is the displacement vector in the time domain. The global mass matrix, stiffness matrix and load matrix are assembled from the element mass matrix $M^e$, stiffness matrix $K^e$ and force matrix $F^e$. The element load matrix is shown in equation 14:
$$F^e = \int (N^e)^T f^e\, dx \quad (13)$$

## 2.2 Numerical Laplace based BSWI FEM in beam structure

As we have developed the BSWI finite element model in time domain shown in Eq. 12, Laplace transform will be applied to convert the time domain equation into a Laplace frequency domain equation in matrix form shown as Eq. (14),

$$K_0(s)u_0(s) = F_0(s) \qquad (14)$$

Where $s = \sigma + iw$ is a complex number, $\sigma$ and $w$ are real numbers, and $K_0(s) = [s^2 M + K]$ is the equivalent stiffness matrix in the Laplace domain, $u_0(s)$ and $F_0(s)$ are the displacement vector and force vector in Laplace domain, respectively.

The BSWI element has a large number of interior nodes. Therefore, it still has a great number of DOFs though the total number DOFs is significantly lower than in other methods. In order to reduce the computational cost, dynamic reduction is proposed to use the displacement of outer DOFs to represent the inner DOFs. Specifically, each BSWI43 element has 11 nodes, 9 nodes of which are interior nodes. The equivalent stiffness matrix could be significantly reduced if we apply the dynamic reduction techniques in order to save computational time.

Suppose the external force only acts on or can be transferred equivalently to the boundary nodal DOFs, Eq.(14) can be re-written based on boundary nodal DOFs and interior nodal DOFs.

$$\begin{Bmatrix} K_{11}(s) & K_{12}(s) \\ K_{21}(s) & K_{22}(s) \end{Bmatrix} \begin{Bmatrix} u_1(s) \\ u_2(s) \end{Bmatrix} = \begin{Bmatrix} F_1(s) \\ 0 \end{Bmatrix} \qquad (15)$$

Here 1 and 2 represent boundary and interior nodal DOFs, respectively. Then the displacement of interior nodal DOFs $u_2(s)$ could be represented by the displacement of boundary nodal DOFs $u_1(s)$ based on the second column of the matrix in Eq. (15), and the new equation with reduced matrix is shown below,

$$\overline{K}(s)u(s) = F(s) \qquad (16)$$

Where $u(s) = u_1(s)$, $F(s) = F_1(s)$, $\overline{K}(s) = K_{11}(s) - K_{12}(s)[K_{22}(s)]^{-1}K_{21}(s)$

Thus, we obtain the displacement $u(s)$ in Laplace domain through the accurate equivalent Laplace domain stiffness matrix.

$$u(s) = [\overline{K}(s)]^{-1}F(s) \qquad (17)$$

The time domain displacement could be obtained by applying the inverse Laplace transform. Laplace transform is a symbol based approach, so it is usually used to find the exact solutions which makes it very difficult to implement in discrete domains. In order to obtain numerical results in an easy way, the numerical Laplace method is used in this paper, because the nodal displacement in the time domain is achieved by substituting the frequency domain solution and inverse Laplace transform, which is shown below,

$$u(t) = \frac{1}{2\pi i}\int_{\sigma-i\infty}^{\sigma+i\infty} u(s)e^{st}\,ds = \frac{1}{2\pi i}\int_{\sigma-i\infty}^{\sigma+i\infty}[K(s)]^{-1}F(s)e^{st}\,ds \qquad (18)$$

Where $s = \sigma + iw$ is a complex number, $\sigma$ and $w$ are real numbers, Eq.(18) could be rewritten as,

$$u(t) = \frac{1}{2\pi i}\int_{\sigma-i\infty}^{\sigma+i\infty}[K(\sigma+iw)]^{-1}F(\sigma+iw)e^{(\sigma+iw)t}\,ds \qquad (19)$$

By applying the substitution rule of variables and changing the integration variable $s$ to be $w$, the following equation could be achieved,

$$u(t) = e^{\sigma t}\frac{1}{2\pi}\int_{-\infty}^{+\infty}[K(\sigma+iw)]^{-1}F(\sigma+iw)e^{iwt}\,dw \qquad (20)$$

Since Fast Fourier Transform is very simple to implement in MATLAB, we here find a way to build a relationship between the Laplace transform and the Fourier transform. By considering $[K(\sigma + iw)]^{-1}F(\sigma + iw)$ as a new function in the transform, Eq. (20) is the Fourier transform of $[K(\sigma + iw)]^{-1}F(\sigma + iw)$ then multiply it by a coefficient $\frac{e^{\sigma t}}{2\pi}$. Similarly, the Laplace transform of the activation force is defined as,

$$F(s) = \int_0^{+\infty} f(t)e^{-st}\,dt \qquad (21)$$

As $s = \sigma + iw$, the force in the frequency domain shown in Eq. 21 can be rewritten as,

$$F(\sigma + iw) = \int_0^{+\infty} f(t)e^{-(\sigma+iw)t}\,dt = \int_0^{+\infty} f(t)e^{-\sigma t}e^{-iwt}\,dt \qquad (22)$$

Thus, the Laplace transform of a force $f(t)$ is considered as the Fourier transform of $f(t)e^{-\sigma t}$ if we take $f(t)e^{-\sigma t}$ as a new term by definition. We can then easily use the fast Fourier transform to get (needs to be checked) the symbol operation of Laplace transform.

## 3. NUMERICAL MODELS
### 3.1 Numerical model of beam structure

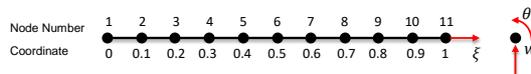

**Fig. 4** The BSWI43 beam element with 2 DOFs

Crack detection in beam structure with a novel Laplace based BSWI FE method

In beam structures, as shown in Fig.4, each node has 2 DOFs, one of which is deflection $w$, and the other one is rotation $\theta$. The displacement functions of both DOFs could be written as a function of the nodal displacement and the shape function

$$w = N^e w^e = \Phi R^e w^e \tag{23}$$
$$\theta = N^e \theta^e = \Phi R^e \theta^e \tag{24}$$

By substituting the displacement equations into the potential energy $U^e$ and kinetic energy $T^e$ in Timoshenko beam, which are functions of the displacement shown below,

$$U^e = \int \frac{EI}{2}\left(-\frac{\partial\theta}{\partial x}\right)^2 dx + \int \frac{GA}{2k}\left(\frac{\partial w}{\partial x} - \theta\right)^2 dx \tag{25}$$

$$T^e = \int \frac{\rho A}{2}\left(\frac{\partial w}{\partial t}\right)^2 + \int \frac{\rho I_y}{2}\left(\frac{\partial \theta}{\partial t}\right)^2 d\xi \tag{26}$$

Where E is Young's modulus, G is the shear modulus, A is the cross sectional area of the rod, k is the shear deformation coefficient, $\rho$ is the density, $I_y$ is the area moment of inertia of the cross-section. By applying Hamilton's variation principle, the stiffness matrix $K^e$ and mass matrix $M^e$ of BSWI rod element are determined as following,

$$K^e = \begin{bmatrix} K_1^e & K_2^e \\ K_3^e & K_4^e \end{bmatrix} \tag{27}$$

$$M^e = \begin{bmatrix} M_1^e & 0 \\ 0 & M_2^e \end{bmatrix} \tag{28}$$

Where, $K_1^e = \frac{GA}{kl_e}\int [R^e]^T [\frac{d\Phi}{d\xi}]^T [\frac{d\Phi}{d\xi}][R^e]d\xi$, $K_3^e = (K_2^e)^T = -\frac{GA}{k}\int [R^e]^T [\frac{d\Phi}{d\xi}]^T [\Phi][R^e]d\xi$

$$K_4^e = \frac{EI}{kl_e}\int [R^e]^T [\frac{d\Phi}{d\xi}]^T [\frac{d\Phi}{d\xi}][R^e]d\xi + \frac{GAl_e}{k}\int [R^e]^T [\Phi]^T [\Phi][R^e]d\xi$$

$M_1^e = \rho A l_e \int [R^e]^T [\Phi]^T \Phi\, R^e d\xi$

$M_2^e = \rho I_y l_e \int [R^e]^T [\Phi]^T \Phi\, R^e d\xi$

### 3.2 Crack model in beam structure

Crack is the most common type of damage in structure. Due to the discontinuity of the displacement at the crack, it is very difficult to construct shape functions. Notching is a simple and common method applied to model crack, while any thickness of the notching is not possible with the characteristics of the crack. The local flexibility method based fracture mechanics is an alternative and effective method, which adopts an equivalent spring model to consider the relationship between the load and the stress concentration at the crack tip.

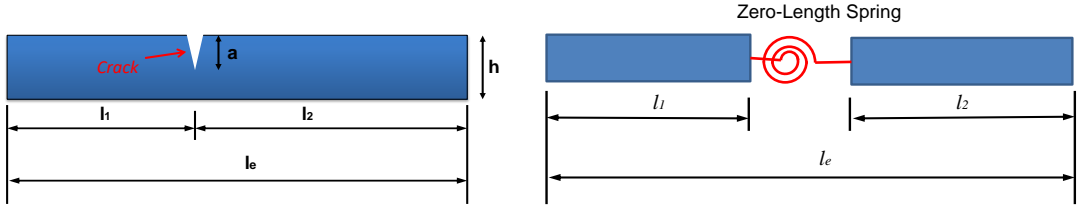

**Fig.5** Crack model of beam structure: 1) cracked element model; 2) spring element model

Due to the influence of axial force, the opening crack mainly occurs in the axial rods. The spring used to simulate the crack in the Timoshenko beam has rotational stiffness and shearing stiffness, where the rotational flexibility $c_b$ and the shearing flexibility $c_s$ can be calculated based on Castigliano's theorem{Przemieniecki, 1985; Tada, 2000}

$$c_b = \frac{72\pi}{Ebh^2}\int_0^a \frac{\alpha}{h^2} f_I^2(\alpha)\, d\alpha \tag{29}$$

$$c_s = \frac{2k^2\pi}{Eb}\int_0^a \frac{\alpha}{h^2} f_{II}^2(\alpha)\, d\alpha \tag{30}$$

Where,

$$f_I(\alpha) = \sqrt{\frac{\tan(\pi\alpha/2h)}{\pi\alpha/2h}} \times \frac{0.752 + 2.02\alpha/h + 0.37[1-\sin(\pi\alpha/2h)]^3}{\cos(\pi\alpha/2h)} \tag{31}$$

$$f_I(\alpha) = \sqrt{\frac{\tan(\pi\alpha/2h)}{\pi\alpha/2h}} \times \frac{0.752 + 2.02\alpha/h + 0.37[1-\sin(\pi\alpha/2h)]^3}{\cos(\pi\alpha/2h)} \tag{32}$$

Here, $\alpha$ is the current location coordinate and $d\alpha$ is the integral infinitesimal, as shown in Fig.5; $E$ is the Young's modulus; $b$ and $h$ are the width and height of cross-section, respectively;

### 4. NUMERICAL EXAMPLES

Several numerical examples of wave propagation simulations in rod structures are proposed to validate the Laplace based wavelet finite element method, a uniform rod is used in the numerical simulation, the geometry parameter and material properties are shown in Table 1.

| Length (mm) | Young's Modulus (GPa) | Poisson's ratio | Density (kg/m3) |
|---|---|---|---|
| 1500 | 200 | 0.3 | 7800 |

**Table 1.** Geometry parameter and material properties of the beam structure

An excitation signal with a 5-cycle sinusoidal tone burst is picked for the wave propagation simulation in the rod structure, the single central frequency is 100kHz. The excitation signal in time domain is listed in Eq. 33.

Also, plots of the excitation signal in time domain and wavelet time-frequency spectrum are shown in Fig. 6.

$$f(t) = \begin{cases} \frac{1}{2}[1 - \cos(40\pi t)]\sin(200\pi t) & 0 \leq t \leq 0.05ms \\ 0 & others \end{cases} \quad (33)$$

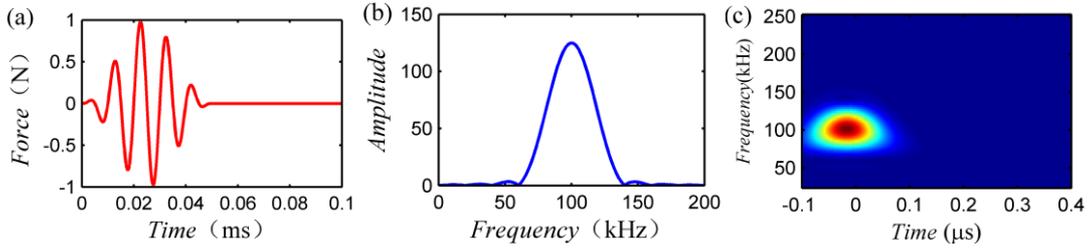

**Fig. 6** Excitation signal : (a) Time-domain diagram; (b) Frequency spectrum; (c) Wavelet time-frequency spectrum

The element size and time step increment are dependent on the wavelength of the excitation signal, which is related to the maximum frequency. The largest frequency of interest for this excitation signal is defined as $f_{max} = 150kHz$. Thus we could get the minimum period for this signal as $T_{min} = 6.67\mu s$ with the shortest wavelength as $\lambda_{min} = 33.75$mm. 20 integration points per period are picked for the simulation, which means that the time interval is $\Delta t = T_{min}/20 = 0.33\mu s$. Since the length of the excitation signal is $t_f = 0.05ms$, the sample points are set as $N = t_f/\Delta t + 1$.

**4.1 Mesh and time dependence study of LWFEM**

As we could see from Fig.6, the Laplace based BSWI method provides very reliable results of ultrasonic wave propagation in rod. We would like to study the advantages and disadvantages of this method. For comparison, conventional FEM can also be applied to compare the advantages and disadvantages between the two methods. From the time frequency analysis of the wave propagation in a rod structure, the central frequency is moving along the rod as the wave propagates along the rod, which also proves the validity of our method.

Firstly, in order to study the influence of the element size on the simulation results, different element sizes with same time intervals for these three methods are studied. The time interval is set as $\Delta t = T_{min}/20 = 0.33\mu s$ to ensure 20 integration time steps per period, while the element size or the number of elements per wavelength (EPW) is different. The displacement response at the middle point of the rod is picked for comparison for these two different numerical methods. Also the EPW is different for various simulation methods. For conventional FEM, the EPW is set as 5, 10, 15, and 20. The simulation result for conventional FEM with different EPWs is shown in Fig. 7. As it can be seen from the plot, the arrival time for different wave packets can be obtained and the group velocity can be calculated and the value is 5053m/s when the EPW is 20, the calculated velocity is very close to the theoretical value $c_0 = \sqrt{E/\rho} = 5063m/s$. In comparison, in the Laplace based wavelet finite element method, we will also set four different values for the number of elements per wavelength, which are set as 0.15, 0.3, 0.45 and 0.6. The results for LWFEM are shown in the right plot in Fig.7. As we could see from this plot, the results converge very quickly when EPW is bigger than 0.3, and 0.45 is a good fit for future simulations. This means that we only need 0.45 element per wavelength in LWFEM to get ideal ultrasonic wave propagation results.

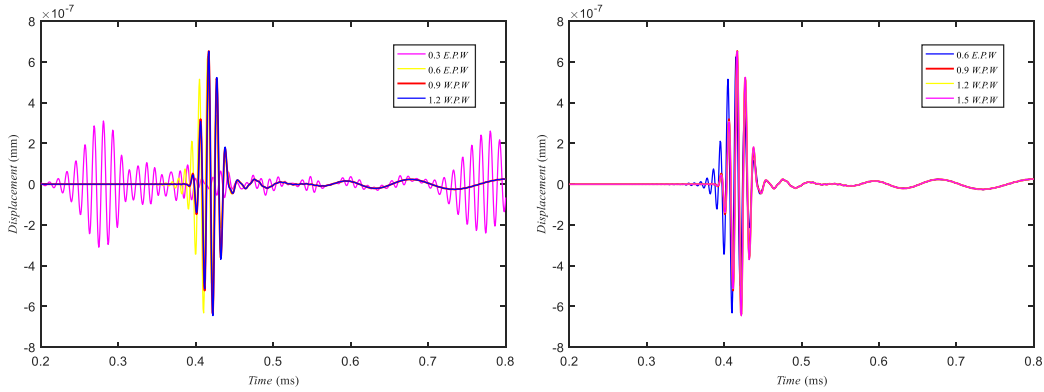

**Fig.7**. Sensitivity study of element size with LWFEM in steel beam

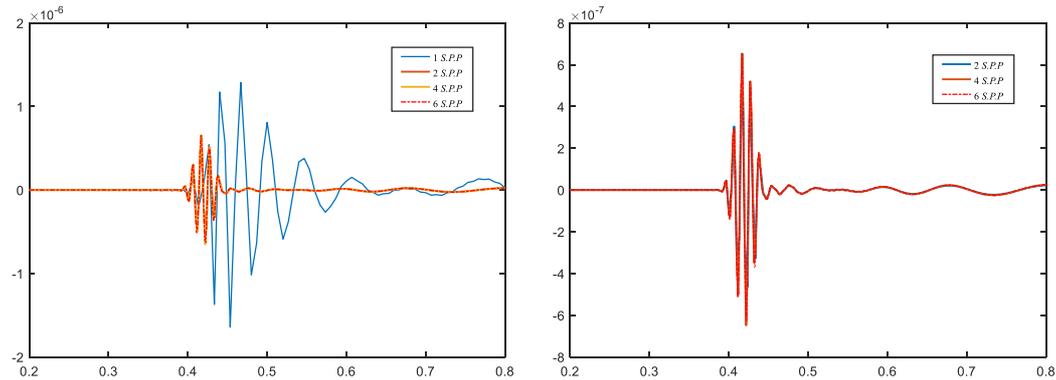

**Fig.8**. Sensitivity study of time interval for LWFEM in steel beam

Another important factor in the finite element method for time integration is the setup of time step, selecting a good value for the time interval is very important since it can influence the accuracy of the results as well as the time cost during computation. Choosing an appropriate time interval value which could both ensure the accuracy of the results and avoid having a long computational time. Here we would like to come up with the concept of a number of integration steps per period (denoted as SPP). For the finite element method, we set SPP as 5, 10, 15, 20, and the SPP is set as 1, 2, 4, 6. The comparison results for the two methods are shown in Fig.8. As shown in Fig.8, the finite element method converges quickly when SPP is larger than 15, but the same value for LWFEM is 2. From the comparison we see that the finite element method needs a smaller time step, while the LWFEM only needs 1/5 of the time interval needed by the finite element method.

**4.2 Wave propagation in Timoshenko beam**

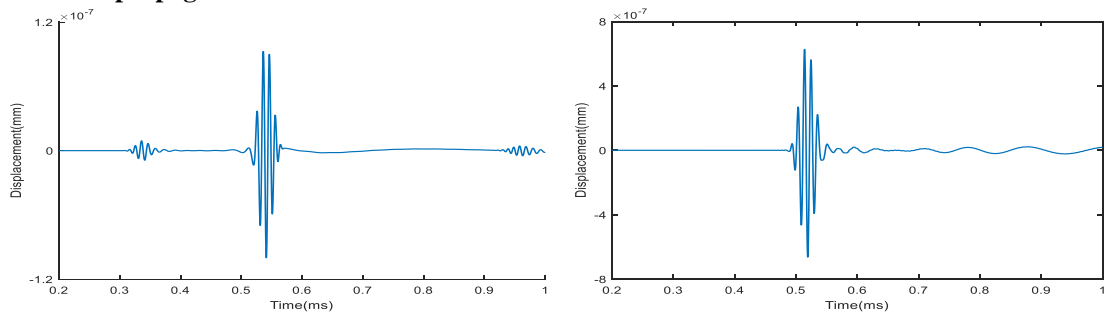

**Fig.9** Deflection in the right end of beam: Left, Aluminum beam; Right, Steel beam

In this part, two beams made with different materials are tested for comparison. The geometry parameters for both beams are the same, and the material properties of the Aluminum beam are shown in Table 2. The specific properties of steels are different because of processing and compositions{Cheng, 2016 & 2017}. For comparison, the time series of deflection at the right end of the two beams is shown in Fig. 9. As shown in Fig.9, the aluminum beam has two more waveforms during this period, the later wave is the reflection wave of the first wave, which means that both additional waveforms originate from the same waveform, namely the second mode. The second mode comes from the excitation frequency, which is higher than the cut-off frequency of this mode in the Aluminum beam{Doyle, 1997}.

| Length | Young's | Poisson's | Density |
| --- | --- | --- | --- |

| (mm) | Modulus (GPa) | ratio | (kg/m3) |
|---|---|---|---|
| 1500 | 200 | 0.3 | 7800 |
| 1500 | 70 | 0.3 | 2730 |

Tab.2 Material properties in Steel and Aluminum beam

Figure 10 shows the waveform of deflection at different times along the beam for both Aluminum and steel beams. The time series picked in Fig.10 are from $40\mu s$ to $320\mu s$ with a time difference of $40\mu s$ for different observations, and the displacement series are normalized for good comparison. As we could see from Fig.10, the wave propagation in steel beam is a little faster than the velocity in Aluminum. The second mode is also observed here and is propagating much faster than the first mode. For a better view, the deflection along the beam at time $t = 120\mu s$ and its time frequency analysis are shown in Fig.11. From Fig.11, we see a clear second mode in Aluminum beam instead of steel beam.

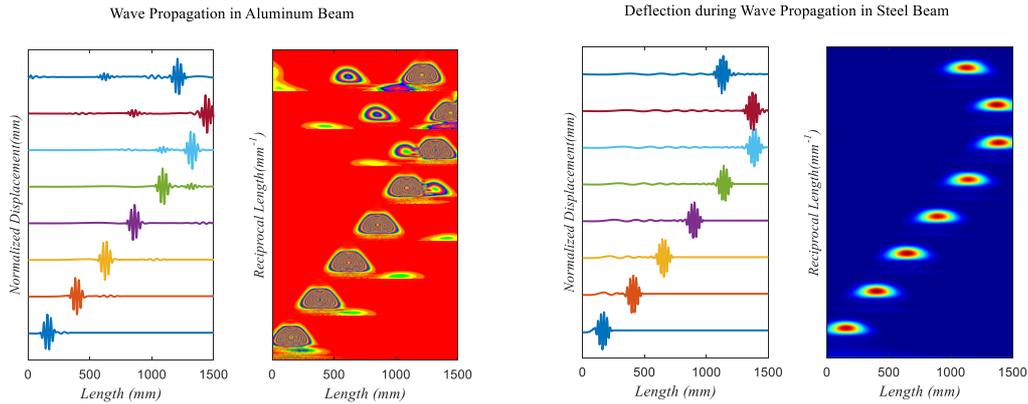

**Fig.10** Waveform of deflection in Timoshenko beam at different time: Left, Aluminum beam; Right, Steel beam

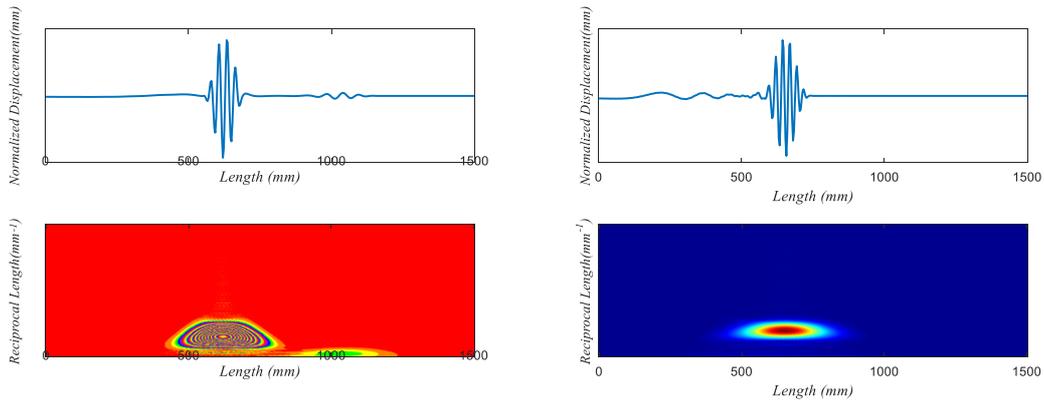

**Fig.11** Time frequency of Deflection in Timoshenko beam at t=120us: Left, Aluminum beam; Right: Steel beam

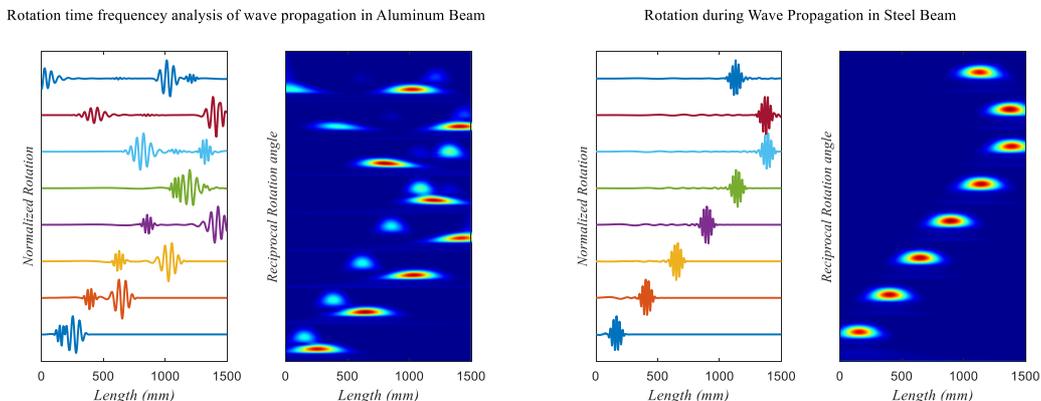

**Fig.12** Waveform of Rotation in Timoshenko Beam at different time: Left, Aluminum beam; Right, Steel beam

### 4.3 The velocity dispersion in beam

The wavelet transform provides more information on ultrasonic wave propagation in a beam structure. Since the wavelet transform is a very good time-frequency analysis tool, we would like to study the time-frequency properties of ultrasonic wave propagation in a rod structure. A new excitation signal is proposed to study the velocity dispersion of guided waves in a rod structure, where double center frequencies 100Khz and 200Khz are included in this excitation signal. Also, the equation of this new excitation signal is shown in Eq.34 and the plots' information is shown in Fig.9:

$$f(t) = \begin{cases} \frac{1}{2}[1 - cos(40\pi t)] \left[\frac{1}{2}sin(200\pi t) + \frac{1}{2}sin(400\pi t)\right] & 0 \leq t \leq 0.05ms \\ 0 & others \end{cases} \quad (34)$$

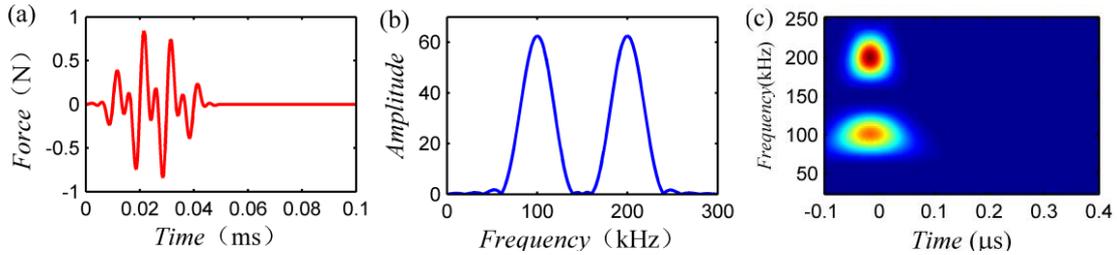

**Fig. 13** Excitation signal for velocity dispersion study: (a) Time-domain diagram; (b) Frequency spectrum; (c) Wavelet time-frequency spectrum

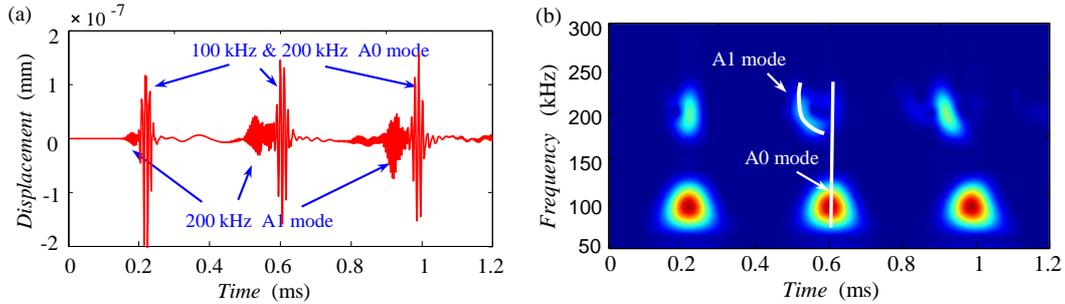

**Fig. 14** The displacement responses at middle point of Timoshenko beam subjected to excitation II: (a) Time-domain diagram; (b) Wavelet time-frequency spectrum

The beam is divided into 36 BSWI Timoshenko beam elements and the displacement responses at the middle point of beam are shown in Fig.14. We can see that the group velocities of transverse waves in beam are slower than that of longitudinal waves in rod. In the vicinity of 100Hz, there is only one mode of wave and the group velocity dispersion is not obvious. However, in the vicinity of 200Hz, two modes of wave are excited and the waves of second mode have obvious dispersion characteristic.

Hence, the development and selection of proper BSWI element are critical for FFT-based BSWI method to simulate wave propagation. It is also an important preparation for SHM to select the proper frequency and mode of wave according to the dispersion property of waves.

### 4.4. Nondestructive testing in a cracked beam structure with LWFEM

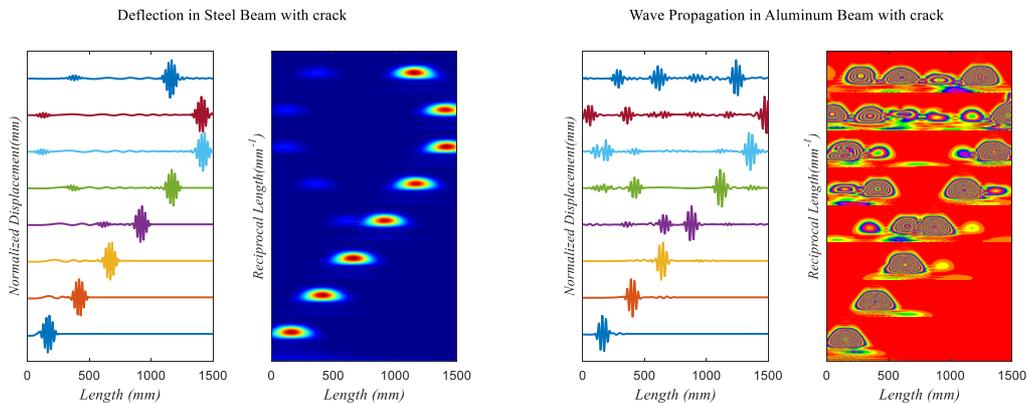

**Fig.15.** Deflection in beam with crack in the middle with Laplace based BSWI method; Left: Steel beam; Right: Al beam

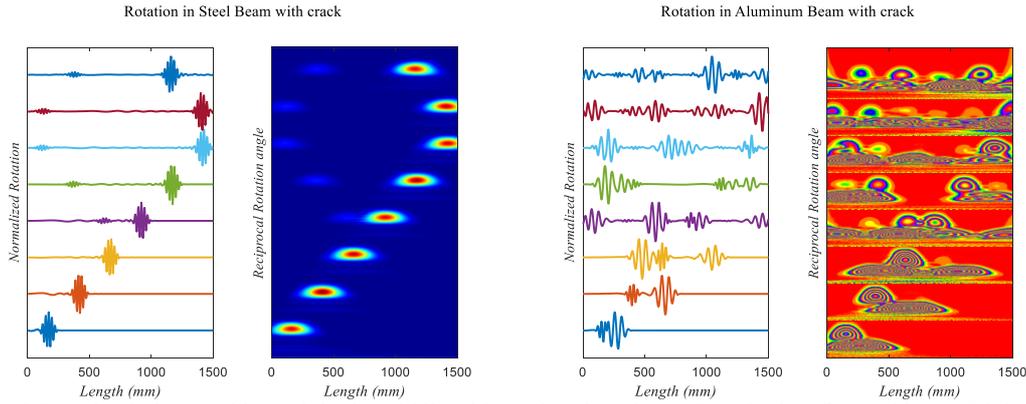

**Fig.16.** Rotation in beam with crack in the middle with Laplace based BSWI method; Left: Steel beam; Right: Al beam

Figure 15 and 16 show the ultrasonic guided wave propagation in the beam with a crack in middle of beam, in which the excitation signal is shown in Eq. 23 and the crack depth is 20% of the width of the rod. The left plot shows the deflection and time frequency analysis in a steel beam at different times with a time difference of $40\mu s$, while the right plot shows the same deflection in a different material, an Aluminum beam.

As we could see from the left plot, the waveform at time $160\mu s$ is different from the waveform at time $120\mu s$, which means that there is a wave that is reflected by the crack. Also as time goes on, the waveform reflected by the crack is more and more obvious, while the excitation wave goes across the crack and propagate along the rod structure until it is reflected by the right end of the rod again. At the same time, the wave reflected by the crack propagates along the rod to the left end and is reflected by the left end of the rod. The time frequency analysis results of the wave propagation signal along the rod structure shows that the central frequency is moving along the rod, and the central frequency is separated into two parts as the wave signal is passing the damage. Also if we take a look at the right plot which shows wave propagation in cracked Aluminum beam, we could clearly see the two wave modes as we analyzed in 4.2, and the two wave modes are reflected by the crack and propagate along the beam.

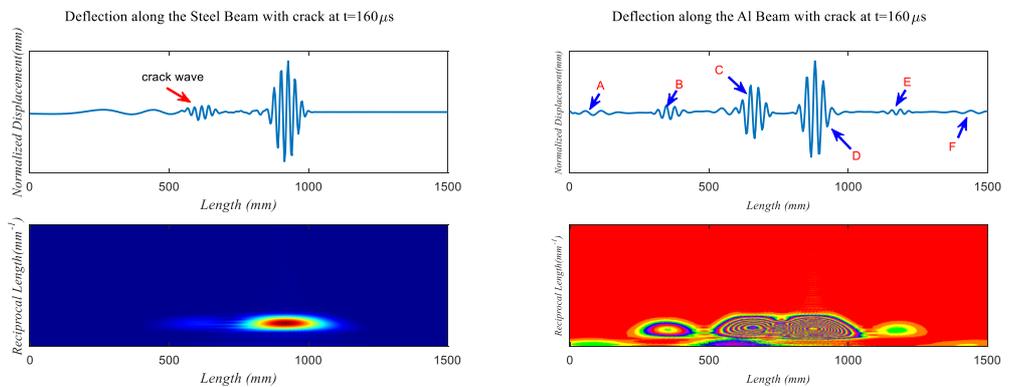

**Fig.17.** Deflection along the beam with crack in the middle with Laplace based BSWI method; Left: Steel beam; Right: Al beam

Figure 17 shows the deflection along the beam with a crack in the middle of the beam at time $t = 160\mu s$, in which the left plot shows the deflection and time frequency analysis in the steel beam while the right plot shows the deflection and time frequency analysis in the Aluminum beam. As shown in the plots, two different modes are obtained in the Aluminum beam while there is only one mode in the steel beam, and two waves are shown in the steel beam, one of which is the directly arrived (check this) wave signal and the other wave is the crack wave signal reflected by the crack. In the right plot, 6 signal waves are obtained, in which signal A is the reflection of the second mode signal, signal B and C are reflected wave signals by the crack while B is the reflection of the second mode and C is the reflection of the first mode, signal D is the first mode signal and wave F is the second mode signal which shows that the second mode signal runs faster than the first mode signal, wave E is the wave caused by crack of the second mode.

## 4.5. Crack depth and location study on wave propagation in cracked beam structure with LWFEM

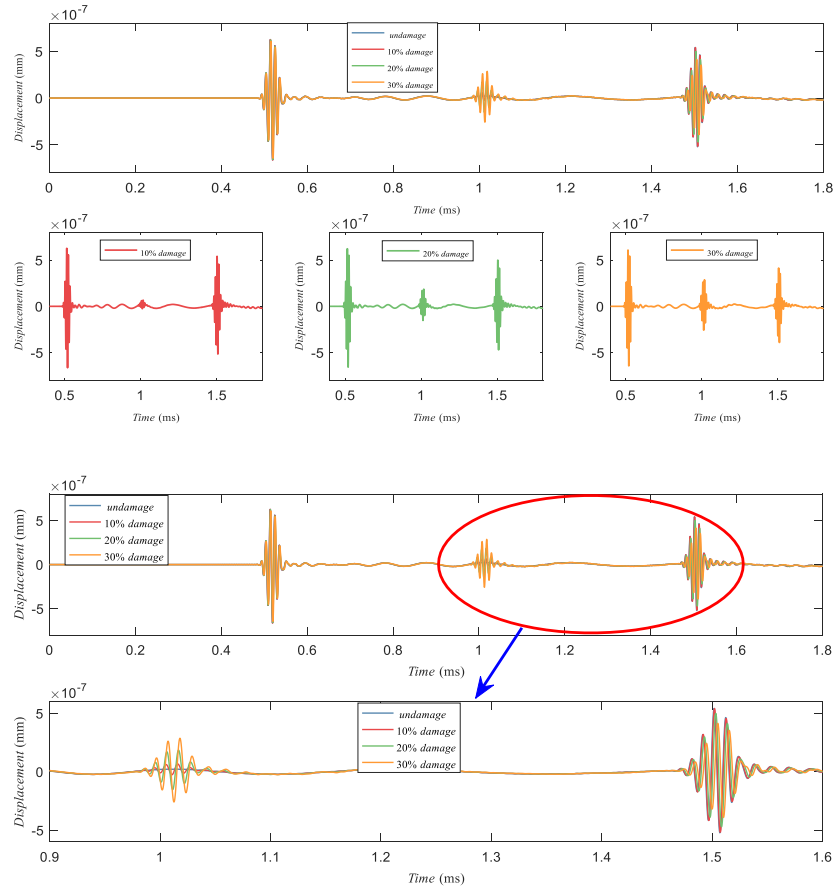

**Fig. 18.** Deflection response on the right end with crack in the middle of beam with different crack depth

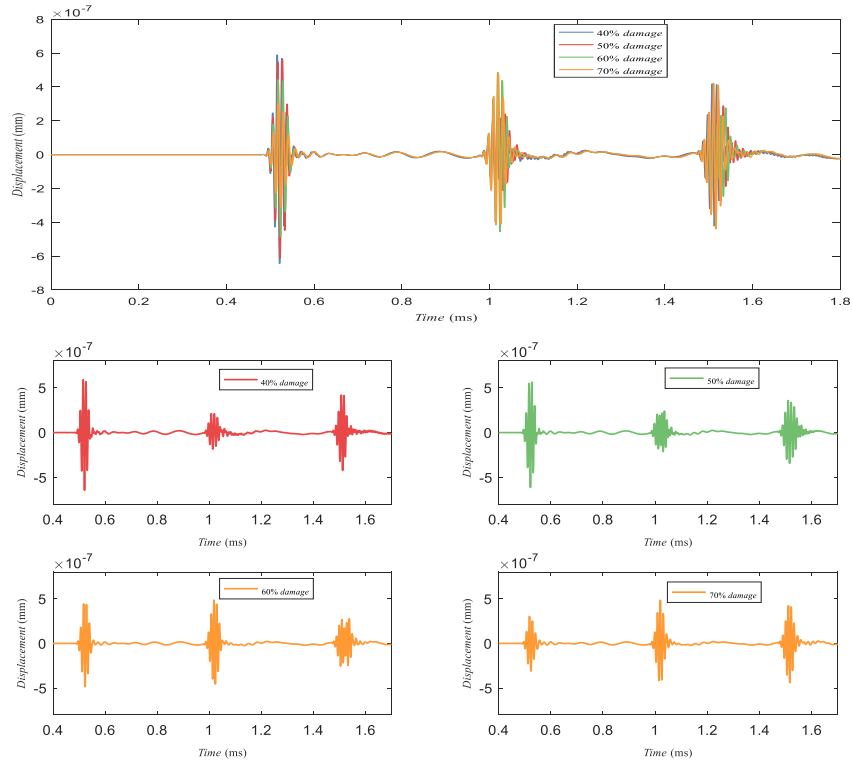

**Fig. 19.** Deflection response on the right end with crack in the middle of beam with different crack depth

Furthermore, we would like to study the influence of the crack depth on wave propagation in the beam; several different ratios of crack depth are compared in our manuscript. Here we fix the crack in the middle of the rod while the crack depths are set with different ratios, Fig.18 and Fig. 19 show the received wave plots at the right end of the beam.

The percentage of damage is evaluated as the ratio of crack depth with respect to the height of beam. As shown in Fig.18, the signal directly received by the right end of the rod structure is almost the same when the damage is small as when the damage ratio is below 30%, but the amplitudes of the flaw signal received by the right end are highly influenced by the crack depth. The amplitude of the flaw signal increases as the crack depth increases. Also, the amplitude of direct wave signal decreases when the crack depth increases as seen in Fig. 19. As shown in Fig. 18 and 19, the amplitude of both direct wave signal and flaw wave signal decreases when the crack depth is increasing.

Another important factor that we studied in this manuscript is the crack location, which is shown in Fig. 20. In this case, we apply the same excitation signal on the left end and receive the signal on the right end of beam. The depth of the cracks are set as $0.2h$, while the locations are set different, one of the cracks is set in the middle of the beam while the other crack is set at the location of $1/4l$. As we see from Fig.14, there are more flaw waves when the crack is located at the $1/4l$ with the same time length. Also, the direct waves received by the right end of rod are the same.

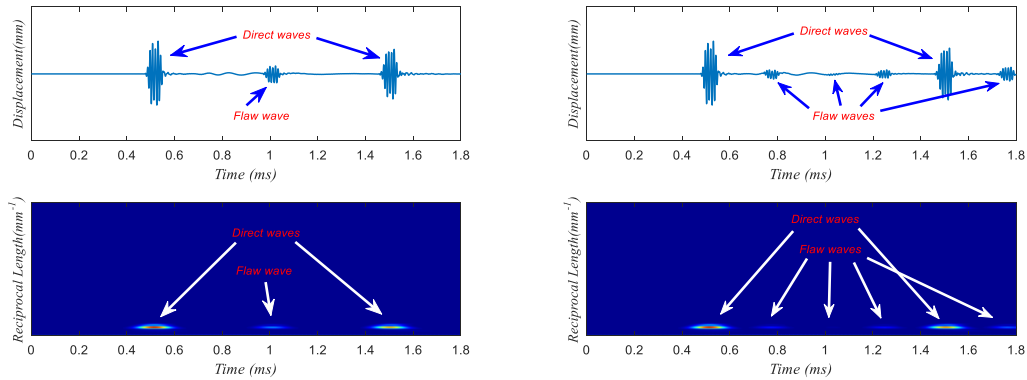

**Fig.20.** Wave signal received by the right end of beam with different crack locations: Left-0.5l; Right-0.25l

## 5. Conclusion

In this manuscript, a novel numerical Laplace based wavelet finite element method is proposed for ultrasonic wave propagation and nondestructive testing in rod structures. Laplace transform is a more advanced method than fast Fourier transform. Laplace transform does not depend on the periodic assumption while Fourier transform does. Also BSWI is a wavelet based finite element method that has been used in ultrasonic wave propagation and has many advantages. By combining the advantages of the two methods, the following conclusions were made:

1. Laplace transform is a symbol-based transform method, but is still applicable via a numerical method through fast Fourier transform, but Laplace transform abandons the periodic assumption of FFT.

2. By comparing the group velocity and wave propagation in beam structure, we see that LWFEM is a very reliable numerical method that can be used in ultrasonic wave propagation and nondestructive testing of beam structures.

3. By studying the sensitivity of the mesh size and time interval with different numerical methods, we conclude that LWFEM has much lower element size and time interval requirements than traditional FEM but can still provide the necessary accuracy of results.

4. LWFEM is a reliable numerical method for nondestructive testing in beam structures and could recognize both small and large damages in the beam structure. The crack location also has a great influence on the received signals on the beam structure.

5. The signal directly received by the right end of the beam structure is almost the same when the damage is small and when the damage ratio is below 30%, but the amplitude of direct wave signal decreases when the crack depth increases. Moreover, the amplitudes of the flaw signal received by the rod are highly influenced by crack depth.

6. The material properties also have a great influence on the wave propagation in beam structures. Different modes can be achieved with different material properties.

**ACKNOWLEDGMENTS**

The authors are grateful for the financial support from National Natural Science Foundation of China (NSFC) under Grant No. 51478079, and the National Fundamental Research Program of China under Grant Nos. 2011CB013703, DUT15LAB11 and Natural Sciences Found of China (No. 51708251).